\documentclass[a4paper,10pt]{article}
\usepackage{graphicx}
\usepackage{amssymb}
\usepackage[utf8]{inputenc}
\usepackage[english]{babel}
\usepackage{xcolor}

\title{Atomic Size Dark Matter Pearls, Electron Signal }
\author{H.B. Nielsen\footnote{Speaker at the  Work Shop
  ``What comes beyond the Standard Models'' in Bled.}, Niels Bohr Institut,
C.D.Froggatt, Glasgow University}
\date{``Bled'', July  2021}

\title{Atomic Size Dark Matter Pearls, Electron Signal}
\author{C. D. Froggatt, Glasgow University\\
H.B.Nielsen, Niels Bohr Institute}

\begin{document}

\maketitle
\begin{center}
  Contributions to Proceedings to the 24th Workshop
  ``What Comes Beyond the Standard Models'', Bled, July 3.-- 11., 2021\\
  H.B. Nielsen gave the talk.

\end{center}
\begin{abstract}
We seek to explain both the seeming
  observation of dark matter by the seasonal variation
  %seeming observation of dark matter by
   of the DAMA-LIBRA data and the observation of
  ``electron recoil'' events at Xenon1T in which the liquid-Xe-scintillator was
  excited by electrons - in excess to the expected background - by the
  {\em same} dark matter model.
   In our model the dark matter consists
  of bubbles of a new type of vacuum containing ordinary atomic matter, say diamond,
  under high pressure ensured by
  the surface tension of the separation  surface (domain wall).
  This atomic matter is surrounded by a cloud of electrons extending
  almost out to atomic size.
  We also seek to explain the self interactions of
  dark matter suggested by astronomical studies of dwarf galaxies and the central structure of
  galaxy clusters. At the same time we consider the interaction with matter in the shielding
  responsible for slowing the dark matter down to a low terminal velocity, so that collisions with nuclei
  in the underground detectors have insufficient energy to be detected.
  Further we explain the ``mysterious'' X-ray line of 3.5 keV from
  our dark matter particles colliding with each other so that the surfaces/skins
  unite. Even the 3.5 keV X-ray radiation from the Tycho supernova remnant is explained as
  our pearls hitting cosmic rays in the remnant.

  What the DAMA-LIBRA and Xenon1T experiments see is supposed to be our dark matter pearls
  excited during their stopping in the shielding or the air. The most
  remarkable support for our type of model is that both these underground
  experiments see events with about 3.5 keV energy, just the energy of the X-ray line.

  We get a good numerical understanding of the fitted cross section over mass
  ratio of self interacting dark matter observed in the study of dwarf galaxies.
  Also the total energy of the dark matter pearls stopped in the shield is
  reasonably matching order of magnitudewise with the absolute observation
  rates of DAMA-LIBRA and Xenon1T, although the proposed explanation of their ratio 
  requires further development.

  It should be stressed that accepting that the different phases of the vacuum
  could be realized inside the Standard Model, our whole scheme could be
  realized inside the Standard Model. So then no new physics is needed for dark
  matter!

\end{abstract}

\section{Introduction}

For a long time we have worked on a dark matter model \cite{Dark1, Dark2, Tunguska, supernova, Corfu2017, Corfu2019, theline, Bled20}, in which the dark
matter consisted of cm-size pearls which were in fact bubbles of a  new
vacuum type surrounded by a skin caused by the surface tension of this
new vacuum. This skin kept a piece of usual atomic matter highly
compressed inside the bubble. In fitting data with this
model the most and almost only successful fit consisted in that we
fitted, with a common parameter, both the overall rate and the very 3.5 keV energy of the X ray line
originally observed in several galaxy clusters, Andromeda and the
Milky Way Center \cite{Bulbul, Boyarsky, Boyarsky2, Bhargava, Sicilian, Foster}
and supposedly coming from dark matter.
But now it turned out that this successful fitting relation between the
3.5 keV energy and the overall rate of the X-ray radiation only depends on
the density of the pearls or equivalently the fermi momentum
or energy of the electrons kept inside the pearls, but not on the absolute
size of the pearls. Thus we could change the model to make the pearl sizes
much smaller, as we shall do in this article, so that they are e.g. now rather of
atomic size. Really we shall let the pearls be of radius
$r_{cloud \; 3.3 MeV}=5*10^{-12}m$. But even such small pearls get stopped
to some extent by the shielding into which they must penetrate to reach
the underground experiments like the DAMA-LIBRA and Xenon experiments looking
for dark matter. Using an astronomical observation based model by
Correa \cite{CAC} especially, we shall construct a rather definite picture
of our pearls from which we estimate that the pearls hitting the
earth actually get stopped presumably in the atmosphere, but if not there then
at least in the earth shielding. The pearls thereby lose so much speed that it becomes
quite understandable that the Xenon-experiments, looking for nuclei
being hit by them and causing scintillation in fluid xenon, will not see
any such events. However the DAMA-LIBRA experiment \cite{DAMA1, DAMA2} would not
distinguish if it is a nucleus that is hit or some energy is released which
causes the scintillator to luminesce. So only the DAMA-LIBRA experiment
%send energy into the scintillation of any sort,
would be able to
get a signal if the dark matter, e.g. our pearls, could be somehow
excited and emit their excitation energy when they pass through the detector.
In our model we shall indeed suggest that the pearls get excited
and emit their energy by electron emission. That would not be
easy to distinguish for DAMA-LIBRA but would still of course come with
seasonal variation\footnote{We note however that the ANAIS experiment
has failed to see an annual modulation with NAI(Tl) scintillators and their
results \cite{ANAIS} are incompatible with the DAMA-LIBRA results at $3.3\sigma$.}
so that it would be observed as dark matter by
DAMA-LIBRA. Whether the emission is via electrons or nuclei would not matter.
But for the xenon-experiments such electron emission was effectively
not counted for a long time, but now rather recently the Xenon1T experiment has
actually observed an excess of ``electron recoil events''. So they
have now in fact seen an electron emission somehow.

We shall see in section \ref{sec7} that both  the excess
of electron recoil events in Xenon1T \cite{Xenon1Texcess} and the events seen by
DAMA-LIBRA \cite{DAMA1, DAMA2} have the energy of each event remarkably enough centering about
the energy value 3.5 keV of the mysterious X-ray line found astronomically!

This coincidence of course strongly suggests that these events
from DAMA-LIBRA and Xenon1T are related to
dark matter particles that can be excited precisely by this energy
3.5 keV.

In our earlier papers \cite{Corfu2017, Corfu2019, theline} we have already connected the excitability of our pearls by
just this energy 3.5 keV and especially the emission of photons
(or here in the present work also electrons) with just this energy with
a gap in the single particle electron spectrum of the pearls
caused by what we call the homolumo-gap effect.

A very serious warning, which needs an explanation in order to rescue our model,
is delivered by the fact that if as we now suggest the Xenon1T electron recoil event excess
is coming from just the same decay of dark matter excitations as the DAMA-Libra
observation, then these two experiments ought a priori to see equally many
events, say per kg. However, DAMA-LIBRA sees 250 times as many events as
Xenon1T sees excess events.

We shall postpone this question to a later article in detail, but the hope
for now is that the Xenon1T experiment has the observed decaying pearls
falling through a fluid, namely the fluid xenon, while the scintillator in
DAMA-LIBRA is a solid made from NaI(Tl). The pearls are likely to
form a little Xe-fluid bubble around them and flow or fall through the
xenon-fluid, while they will much more easily get caught so as to almost sit still
or only move much slower through the NaI scintillator. If so the pearls
with their supposed excitations would spend much more time in the
DAMA-LIBRA NaI than in a corresponding volume of xenon-liquid.

In the following section \ref{Pearl} we describe how the particles making up the
dark matter in our model are imagined to be bubbles of the size $R=
r_{cloud \; 3.3MeV}=5*10^{-12}m$ with heavy atomic matter inside, which is surrounded out
to a radius $r_{cloud \; 3.5 keV}= 5*10^{-11}m$ by electrons. Here the quantities 3.3 MeV and
3.5 keV in the subscripts are the
numerical electric potentials felt by an electron at the distances mentioned.
A special point to note in this section already present in the earlier articles
about the big pearls is the homolumo-gap effect, causing a band or gap in the energy
levels without any single particle electron eigenstates.  The width of this gap is fitted to
the 3.5 keV line in the observed X-ray spectrum from galaxy clusters, the
Milky Way Center etc. \cite{Bulbul, Boyarsky, Boyarsky2}.

Next in section \ref{ngi} we briefly review astronomical
observations and modelling of the dark matter, which suggests
the idea that dark matter interacts with itself (strongly interacting dark matter SIDM).
It is only when the corresponding cross section $\sigma$
is divided by the
particle mass $M$ do we have a combination that has any chance of being observed
by its effects on the atomic matter. In fact this ratio $\frac{\sigma}{M}$ matches well with the atomic physics
structure of our pearls including the cloud of electrons outside the bubble
itself.

In  section \ref{Achievements} we list a series of numerical successes of our
model for the dark matter, hopefully making the reader see that there is really
some reason for it being at least in some respects correct.

In section \ref{Impact} we restress that our dark matter pearls get stopped and
at the same time excited, mainly to emit quanta of energy 3.5 keV, in the air
and/or in the shielding above the experiments. According to our best estimates
they get stopped already about 53 km up in the air. It is the braking energy from this
slowing down that is supposed to feed the excitations.

A special estimation, based on energy considerations, of whether the number of events seen by
DAMA-LIBRA and by the Xenon1T electron recoil excess are of a
reasonable order of magnitude is put forward in section \ref{sec5}.
The success of such an estimation has to be rather limited in as far as the rates of the two
observations - that should have been the same if we do not include the possibility of
faster or slower motion through the detectors - deviate by a factor of 250.

In section \ref{sec7} we call attention to the perhaps most remarkable
fact supporting a major aspect of our model: That the energy per event
for both DAMA-LIBRA and the Xenon1T-electron recoil excess centers around
3.5 keV, just the energy of the photons in the mysterious X-ray line
seen in galactic clusters mentioned above! So all three effects should
correspond to the emission of an electron or photon due to the same energy
transition inside dark matter.

Finally in section \ref{Conclusion} we conclude and provide a short outlook.

\section{Pearl}
\label{Pearl}
%\begin{frame}

  {\bf Dark Matter Atomic Size Pearls, Electronic 3.5 keV Signal }

\begin{figure}
\includegraphics[scale=0.9]{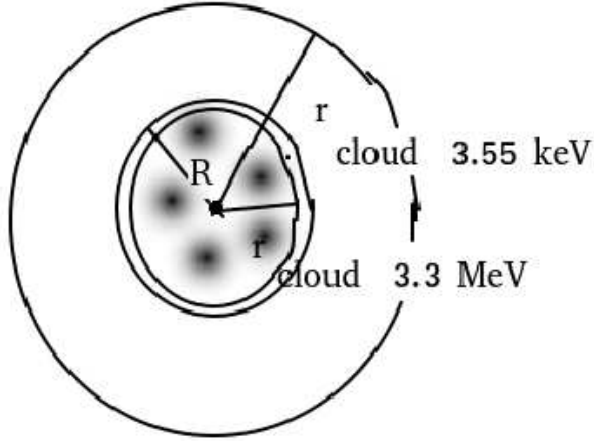}
%\end{frame}
%\begin{frame}
\caption  {\label{Pearl1} The figure illustrates the bit smaller than atom-size
    complicated/macroscopic dark matter particle in our
    model, a pearl.}
\end{figure}

We sketch the structure of our small dark matter pearls in Figure 1.
  \begin{itemize}
  \item In the middle is a spherical bubble of radius
    \begin{eqnarray}
      R \approx r_{cloud \; 3.3 MeV}&\approx & 5*10^{-12}m.
     % \hbox{where } r_{cloud \; 3.3 MeV}&=&
      %\hbox{``where potential for electron is $3.3 MeV$''}\nonumber
    \end{eqnarray}
   % \vspace{-2mm}
    Here $r_{cloud \; 3.3 MeV}$ denotes the radius where the electron potential
    is 3.3 MeV, which is identified with the Fermi energy $E_f$ of the
    electrons in the bulk of the pearl - i.e. inside the radius $R$. We
    estimated the value $E_f = 3.3$ MeV in previous papers \cite{Corfu2019, theline, Bled20} by fitting the
    overall rate of the intensity of the 3.5 keV line emitted by galactic
    clusters and the very frequency 3.5 keV of the radiation in our model.
  \item The outer radius
  \begin{equation}
  r_{cloud \; 3.5 keV} \approx 5 *10^{-11} m
  \end{equation}
  is where the electron potential
    is 3.5 keV.  By our story of the ``homolumo gap'':  the
    electron density crudely goes to zero at this radius. (It gradually falls
    in the range between $r_{cloud \; 3.3 MeV} $ and
    $r_{cloud \; 3.5 keV}$).
    \end{itemize}
%\end{frame}
%\begin{frame}

  {\bf The electron density and potential in the pearls}

\begin{figure}
	  \includegraphics[scale=0.8]{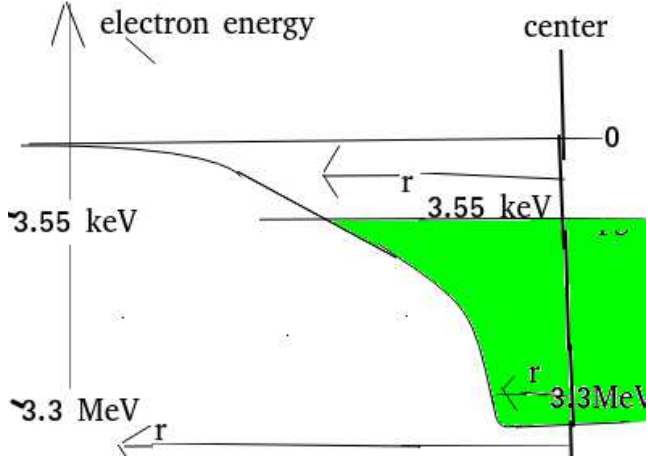}
%\end{frame}
%\begin{frame}
          \caption{\label{density} Explaining the electron density and electric potential in the Pearl}
\end{figure}

  \begin{itemize}
  \item Due to an effect, we call the {\em homolumo-gap effect} \cite{Corfu2017, jahn}, the nuclei
    in the
    bubble region and the electrons themselves become arranged in such a way
    as to prevent there from being any levels in an interval of width
    3.5 keV. So, as illustrated in Figure \ref{density},
    outside the distance $r_{3.5 keV} = r_{cloud \; 3.5 keV}$ from
    the center of the pearl at which the Coulomb potential is $\sim$ 3.5 keV
    deep there are essentially ($\sim$  in the Thomas-Fermi approximation) no
    more electrons in the pearl-object.
  \item The radius $r_{3.3 MeV}= r_{cloud \; 3.3 MeV}$ at which the potential
    felt by an electron is 3.3 MeV deep, is  supposed to be just the radius
    to which the many nuclei inside the pearl (which replace
    the single nucleus in ordinary atoms) reach out. So inside the bubble
    the potential is much more flat.
  \end{itemize}
%\end{frame}
%\begin{frame}

%  {\bf Explanation of Electron Density and Potential Figure Continued.}
  \begin{itemize}
  \item The energy difference between the zero energy line and the effective
    Fermi surface, above which there are no more electrons, is of order
    3.5 keV, the energy so crucial in our work.

  \item Since in the Thomas-Fermi approximation there are no electrons outside
    roughly the
    radius $r_{3.5 keV}=r_{cloud \; 3.5 keV}$, this radius will give the maximal
    cross section, even for very low velocity $\sigma_{v\rightarrow 0}$.
    \end{itemize}
%\end{frame}
%\begin{frame}

  {\bf The homolumo gap effect.}

\begin{figure}
    \includegraphics{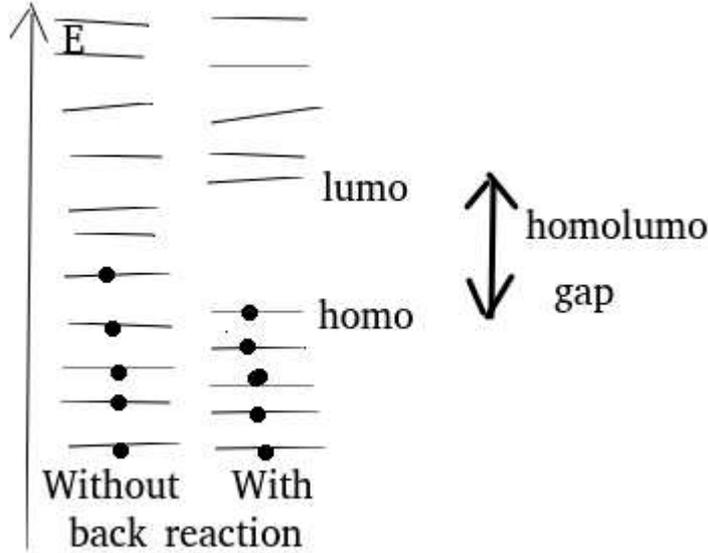}
 %   \end{frame}
%\begin{frame}
\caption{\label{homolumo} Explanation of Homolumo-gap effect}
\end{figure}
        Let us consider the spectrum of energy levels for the electrons
        in a piece of material, e.g. one of our pearls, and at first assume
        that the positions or distributions of the charged particles in the material
        are fixed.

        Then the ground state is just a state built e.g. as a Slater
        determinant for the electrons being in the lowest single electron
        states, so many as are needed to have the right number of electrons.

        But now, if the charged particles can be moved due to their interactions, the
        ground state energy could be lowered by moving them so that the filled
        electron state levels get lowered.

        {\em So we expect introducing such a ``back reaction'' will lower
          the filled states.}
 %     \end{frame}
%\begin{frame}

  %      {\bf Why the name ``homolumo''}

        When the filled levels get moved downwards,
         then the homo = ``{\bf h}ighest {\bf o}ccupied {\bf m}olecular
        {\bf o}rbit'' level will be lowered and its distance to the
        next level, the
        lumo (= {\bf l}owest {\bf u}noccupied {\bf m}olecular {\bf o}rbit),
        will appear extended on the energy axis.
 %     \end{frame}
%\begin{frame}

        {\em We believe that we can estimate the homolumo-gap $E_H$.}

        Using the Thomas-Fermi approximation - or crudely just some dimensional
        argument where the fine structure constant has the dimension of velocity -
        we calculated the homolumo gap in highly compressed ordinary matter for
        relativistic electrons:
        \begin{eqnarray}
          E_H &\sim& (\frac{\alpha}{c})^{3/2}\sqrt{2}p_f\\
          \hbox{where } \quad p_f &=& \hbox{Fermi momentum} \\
          \frac{\alpha}{c}&=& \frac{1}{137.03...}
        \end{eqnarray}
        %\vspace{-2mm}
        (the $\sqrt{2}$ comes from our Thomas-Fermi calculation).

        It is by requiring this homolumo-gap to be the 3.5 keV energy
        of the X-ray line mysteriously observed by satellites from clusters
        of galaxies, Andromeda and the Milky Way Center that we estimate
        the Fermi-energy to be $E_f \approx p_f =$ 3.3 MeV
        in the interior bulk of the pearl.

        %\end{frame}
%\begin{frame}

  {\bf Brief summary of theoretical ideas underlying our dark matter pearls}
%  {\bf Dark Matter Pearls, Sizes Fitted to
 %   $\frac{\sigma}{M}|_{v\rightarrow 0}=150cm^2/g$}

  \begin{figure}
  \includegraphics[scale=0.9]{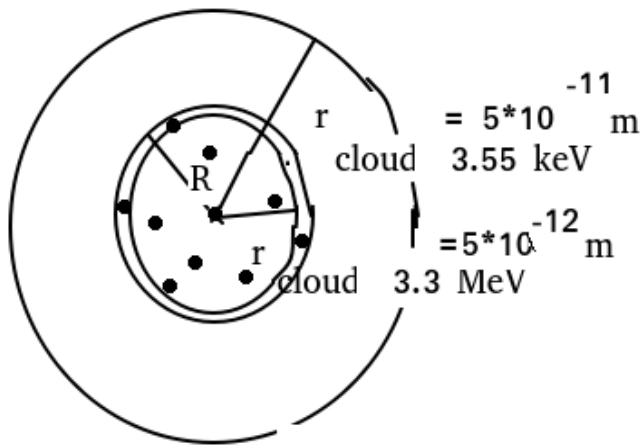}
%\end{frame}
%\begin{frame}
\caption{ \label{Pearl2}Our Picture of Dark Matter Pearls.}
\end{figure}

\begin{itemize}
\item{\color{blue} Principle } Nothing but Standard Model!
  { (Seriously it would mean not in a BSM-workshop.)}
\item{\color{blue} New Assumption} Several Phases of Vacuum with Same Energy
  Density; this is the so-called Multiple Point Principle \cite{Corfu2017, Corfu2019, MPP1, MPP2, MPP3, MPP4, tophiggs, Corfu1995}.
\item{\color{blue} Central Part} Bubble of New Phase of Vacuum with e.g. carbon
  under very high pressure, surrounded by a surface with tension $S$
  (= domain wall) providing the pressure.
\item{\color{blue} Outer part} Cloud of Electrons much like an ordinary atom
  having a nucleus with a charge of order ten to a hundred thousand ($Z \approx 5*10^4$ effectively).
\end{itemize}
%\end{frame}]

\section{Non-gravitational Interactions}
\label{ngi}
The collisionless cold dark matter model provides a good description of the
large scale structure of the Universe. However there are various problems
at small scales \cite{annika, SIDM} for the hypothesis that dark matter only has
gravitational interactions. Originally Spergel et al \cite{firstSIDM}
suggested that the lack of a peak or cusp in the center of galaxy clusters, as
expected for cold dark matter with purely gravitational interactions, required
self interacting dark matter with a relatively large cross section. The
relevant parameter is in fact the cross section per mass $\frac{\sigma}{M}$
and for the cores in galaxy clusters, where the collision velocity is
$v \sim$ 1000 km/s, a value $\frac{\sigma}{M} \sim$ 0.1 $cm^2/g$ is needed. The
self interaction can of course be velocity dependent and the cores in spiral
galaxies where $v \sim$ 100 km/s require  $\frac{\sigma}{M} \sim$ 1 $cm^2/g$. In
dwarf galaxies around our Milky Way, where dark matter moves more slowly
$v \sim$ 30 km/s, larger cross section to mass ratios
$\frac{\sigma}{M} \sim$ 50 $cm^2/g$ are needed.

  \begin{figure}
  \includegraphics[scale=0.9]{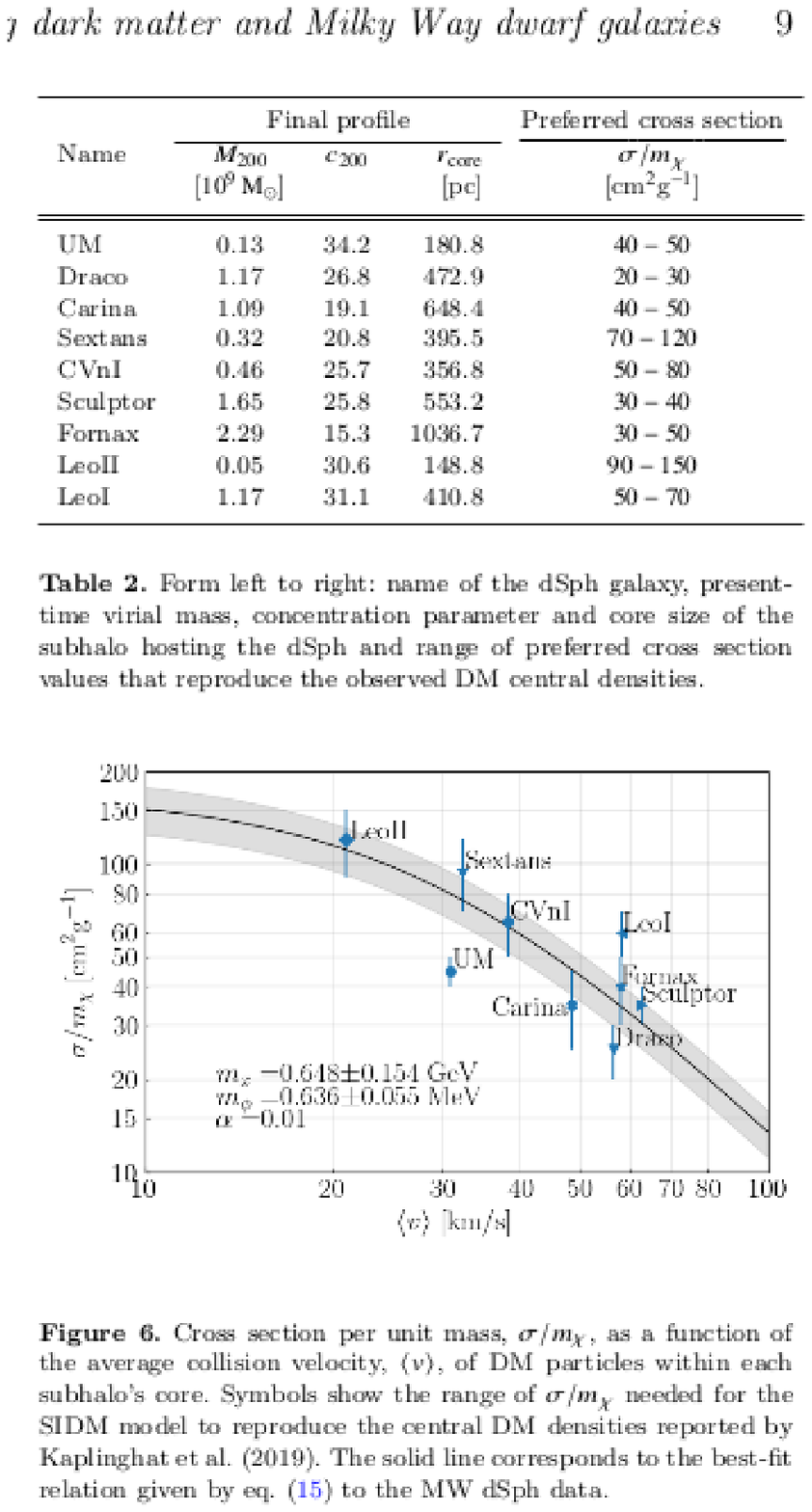}
  \caption{\label{Correa} Cross section per mass $\frac{\sigma}{M}$ of self
    interacting dark matter particles as a function of the collision velocity
    $v$ in dwarf galaxies from reference \cite{CAC}.}
\end{figure}

  Recently Correa \cite{CAC} made a study of the velocity dependence of self
  interacting dark matter. In particular she analysed the Milky Way dwarf
  galaxies and her results are displayed in Figure \ref{Correa}. The
  extrapolation of Correa's fit to the data towards zero velocity points to
  the ratio $\frac{\sigma}{M} \rightarrow$ 150 $cm^2/g$. This ratio can be
  taken as an experimental estimate of the impact area over the mass as seen
  for very soft collisions. In our model the cross section in this low
  velocity limit is given by the extent out to which the electrons
  surrounding our pearls reach. This range of extension of electrons is
  supposed to be given by the requirement that the electron binding energy
  is of the order of the homolumo gap value 3.5 keV. So we denote this
  radius by $r_{cloud \; 3.5 keV}$. Similarly the radius of the bubble
  containing the nucleons inside our dark matter pearl corresponds to a
  radius  $r_{cloud \; 3.3 MeV}$ at which the potential for the electron is
  -3.3 MeV (= Fermi energy of the electrons). The high velocity hard
  collisions of our pearls, supposed to result in the unification of two pearls into a single pearl,
  correspond to interactions between the bubble
  skins with a cross section of order $\pi r_{cloud \; 3.3 MeV}^2$.

  We will now consider the electric potential for our pearl using the
  Thomas-Fermi approximation for a heavy atom \cite{Thomas, Fermi, Spruch}.
  In this approximation the Coulomb
  potential of the ``nuclear" charge Z is multiplied by the Thomas-Fermi
  screening function $\chi(r/b)$ where
 \begin{equation}
 b = 0.88 \frac{a_0}{Z^{1/3}}
 \end{equation}
 and $a_0$ is the Bohr radius.
 The skin of the bubble or ``nucleus" of the pearl mainly acts on the nucleons
 or rather nuclei. So the electrons spread out and an appreciable part, say
 half of them, are outside the central part of the pearl inside the skin.
 Therefore the effective charge $Z$ of the central part of the pearl or
 bubble of the new phase is  e.g. one half of the number of protons inside
 the skin. Assuming also that there are about equally many neutrons and
 protons
 inside the central part, the mass of the pearl is then given order of
 magnitudewise by $M = 4 m_N*Z$, where $m_N$ is the nucleon mass.

 In the Thomas-Fermi approach we are then led to the following equations
 for $r_{cloud \; 3.5 keV}$ and $r_{cloud \; 3.3 MeV}$:
\begin{eqnarray}
  %\pi *r_{cloud \; 3.5 keV}^2/M &=& 150 cm^2/g\\
  \frac{\alpha *Z}{r_{cloud  \; 3.5 keV}}*\chi(r_{cloud \; 3.5 keV}/b)&=& 3.5 \; keV
  \label{3.5keV}\\
   \frac{\alpha *Z}{r_{cloud  \; 3.3 MeV}}*\chi(r_{cloud \; 3.3 MeV}/b)&=& 3.3 \; MeV
  \label{355keV}\\
  b&=& 0.88*\frac{a_0}{Z^{1/3}}\label{bdef}\\
  %M&\ge& 2m_N*Z.
\end{eqnarray}
We identify $r_{cloud  \; 3.5 keV}$ with the radius of the electron
cloud and $r_{cloud \; 3.3 MeV}$ with the skin radius $R$ of the pearl.

It is going to be an important success of our model that we get a similar
value for $R \approx r_{cloud \; 3.3 MeV}$ using another
method to calculate it. We shall use
\begin{equation}
\frac{\sigma}{M}|_{v \rightarrow 0} = 150 \; cm^2/g
\end{equation}
and
\begin{equation}
\sigma = \pi * r_{cloud \; 3.5 keV}^2
\end{equation}
to determine the mass M. Then using the formula for the mass of a pearl in
terms of the radius $R$ and the Fermi momentum \cite{theline, Bled20}
\begin{equation}
\frac{M}{m_N}=\frac{8}{9\pi} * (R*p_f)^3,\label{Mpf}
\end{equation}
we can calculate another value for $R$.

In our updated contribution to the Bled Proceedings from last year \cite{Bled20} we estimated a pearl
mass of $M \sim 10^5$ GeV.
So we take here $Z=5.3*10^4$ as a typical charge in the central part of the
pearl, for which then $b = 1.24 * 10^{-12} m$. Using numerical values for the Thomas-Fermi screening function in the
paper \cite{Parand}, we obtain from (\ref{3.5keV}) the radius of the
electron cloud to be
\begin{equation}
r_{cloud \; 3.5 keV}=
%3.32 *10^{-11}m$
4.96*10^{-11}m \label{3.5k}
\end{equation}
Then assuming the low velocity ratio $\frac{\sigma}{M}=150 \; cm^2/g$ we obtain
\begin{eqnarray}
  M&=& \frac{\pi *(
    %3.32
    4.96*10^{-11}m)^2}{150 \; cm^2/g}\\
  &=&5.2*10^{-19}g\\
  %2.3*10^{-19}g\\
  &=&3.1 *10^5  \; m_N
  %1.3*10^5 m_N
  \end{eqnarray}

As a side remark notice that, using our proposed rule of taking
$Z$ to be a quarter of the number $M/m_N$, we would get
$Z=8*10^4$ to be compared with our input here $5.3*10^4$, which is very well
consistent within a factor 2.

Next using (\ref{Mpf}) with $p_f =3.3 \; MeV = 1.6 *10^{13} m^{-1}$

\begin{eqnarray}
  (R*p_f)^3 &=&
  %2550
  %1.3
  3.1*10^5*\frac{9\pi}{8}\\
   &=&
  %9000
  %4.9*10^5
  10.9*10^5\\
  R * 1.6 *10^{13}m^{-1}&=& \sqrt[3]{
    %9000
    %4.9*10^5
  10.9*10^5}
  %&=&
  %21
  %79
  = 102
  \end{eqnarray}
  giving
  \begin{eqnarray}
   R &=&\frac{
    %79
  102}{1.6*10^{13}m^{-1}}\\
  &=&6.4*10^{-12}m.\label{r3.3M}
  \end{eqnarray}
This is to be compared with the Thomas-Fermi value obtained from
(\ref{355keV}) using the numerical values for $\chi(r/b)$ in \cite{Parand}
  \begin{eqnarray}
   R = r_{cloud \; 3.3 MeV} &=&
  %3.08
  %2.95*1.24 *10^{-12}m\\
  %&=&
  %3.82
  3.66*10^{-12}m. \label{r3.3TF}
  \end{eqnarray}
  These two different estimates of the radius $r_{cloud \; 3.3MeV}$ at which the
  potential is 3.3 MeV essentially coincide to the accuracy of our
  calculation; they deviate by a factor of order unity 6.4/3.7 =1.7. So we
  could claim that formally our model is able to predict the low velocity
  limit $\frac{\sigma}{M}|_{v \rightarrow 0}$ in agreement with the value $150 \; cm^2/g$
  estimated from the study of dwarf galaxies around the Milky Way.

  We shall take the average of the two values  (\ref{r3.3M}) and (\ref{r3.3TF}) as
  our best estimate of the bubble skin radius:
  \begin{equation}
  r_{cloud \; 3.3MeV} = 5.0 *10^{-12} m
  \end{equation}
  and from (\ref{3.5k}) we have the radius of the electron cloud
  \begin{equation}
  r_{cloud \; 3.5keV} = 5.0 *10^{-11} m.
  \end{equation}
  We note that these two radii differ by an order of magnitude, which means
  that the quantity $\frac{\sigma}{M}$ for our pearls should differ by two
  orders of magnitude between low velocities and high velocities, as
  astronomical observations indicate is the case for self interacting dark
  matter \cite{CAC}.

\section{Achievements}
\label{Achievements}
%\begin{frame}

%  {\bf Main Achievements (first), Fitting Low velocity Interaction. }

\begin{itemize}
\item{{\bf \color{blue} Low velocity $\frac{\sigma}{M}|_{v\rightarrow 0}$ cross
    section to mass ratio.}} The a priori story, that dark matter has only
  gravitational interactions seems not to work perfectly: Especially in dwarf
  galaxies (around our Milky Way) where dark matter moves relatively slowly an appreciable self interaction
  cross section to mass ratio $\frac{\sigma}{M}$ is needed.
  According to the fits in \cite{CAC} this ratio has the low velocity limit
  $\frac{\sigma}{M}|_{v\rightarrow 0} = 150 \;cm^2/g$.
  We may say our pearl-model
  ``predicts'' this ratio in order of magnitude.
%\end{itemize}
%\end{frame}
%\begin{frame}

%\end{frame}
%\begin{frame}

%  \includegraphics[scale=0.4]{SIDMvv200702958.eps}

 % \end{frame}
%\begin{frame}

 % \includegraphics[scale=0.9]{fig3200612515.eps}

 % \end{frame}
%\begin{frame}

 % \includegraphics[scale=0.5]{SIDMSsmpfirstremoval68.eps}

%\end{frame}
%\begin{frame}
 % \begin{figure}
  %	\label{Correa2}
  %\includegraphics[scale=0.8]{SIDMvv200702958cropped.eps}
%  \caption{}
 % \end{figure}

 % \end{frame}
%\begin{frame}

 % {\bf Main Achievements (continued), Resolving Disagreement Seemingly
  %  of DAMA and Xenon-experiments.}

%\begin{itemize}
\item{{\bf \color{blue} Can make the Dark Matter Underground Searches get
    Electron Recoil Events}} Most underground experiments are designed to look
  for dark  matter particles hitting the nuclei in the experimental apparatus,
  which is then scintillating so that such hits presumed to be on nuclei can
  be seen.
  But our pearls are excited in such a way that they send out energetic
  {\em electrons} (rather than nuclei) and this does not match with what is
  looked for, except in the DAMA-LIBRA experiment. In this experiment the only signal for
  events coming from dark matter is a seasonal variation due to the Earth
  running towards or away from the dark matter flow.

%\end{itemize}
%\end{frame}
%\begin{frame}

 % {\bf Main Acievement (yet continued), The 3.5 keV X-ray Radiation}

 % \begin{itemize}
  \item{\bf \color{blue} The Intensity of 3.5 keV X-rays from Clusters etc.}
    We fit the very photon-energy 3.5 keV and the overall intensity from a
    series of clusters, a galaxy, and the Milky Way  Center \cite{Bled20} with one parameter
    $\frac{\xi_{fS}^{1/4}}{\Delta V} = 0.6 \; MeV^{-1}$.

  \item{\bf \color{blue} 3.5 keV Radiation from the Tycho Supernova
    Remnant.} Jeltema and Profumo \cite{Jeltema} discovered the 3.5 KeV
    X-ray radiation coming
    from the remnant of Tycho Brahe's supernova, which was unexpected for
    such a small source. We have a
    scenario giving the correct order of magnitude for the observed intensity
    in our pearl model: supposedly our pearls are getting excited by the
    high intensity of cosmic rays in the supernova remnant \cite{Bled20}.
  %\end{itemize}

%\end{frame}
%\begin{frame}

   Even though we can use only the one parameter
    $\frac{\xi_{fS}^{1/4}}{\Delta V}=\frac{2}{p_f}$, it is nice to know the
    notation:

  \begin{eqnarray}
    \Delta V &=& \hbox{`` difference in potential for a nucleon between the}\nonumber\\
    && \hbox{inside
      and the outside of the central part of the pearl''}\nonumber\\
    &\approx & 2.5 \; MeV\\
    \xi_{fS}&=& \frac{R}{R_{crit}} \quad \hbox{estimated to be } \approx 5\\
    \hbox{where } R &=& \hbox{``actual radius of the new vacuum part''}
    \nonumber\\
    &\approx & r_{cloud \; 3.3 MeV}\\
    \hbox{ and } R_{crit}&=& \hbox{`` Radius when pressure is so high}\nonumber\\
    && \hbox{ that nucleons
      are just about being spit out''}
    \end{eqnarray}
 % \end{frame}
The subscript $fS$ on the parameter $\xi_{fS}$ indicates that the surface tension $S$
is fixed independent of the radius $R$.
%\begin{frame}

  %{\bf Main calculational Achievements (yet yet contiued), Energy Consideration}

  %\begin{itemize}
  \item{\color{blue}DAMA-rate} Estimating observation rate of DAMA-LIBRA
    from kinetic
    energy of the incoming dark matter as known from astronomy.
  \item{\color{blue}Xenon1T Electron recoil rate} Same for the
    {\em electron recoil
    excess} observed by the Xenon1T experiment.
  \end{itemize}
In order to explain these last calculational estimates it is necessary to
know how we imagine
the dark matter to interact and get slowed down in the air and the earth
shielding; also how the dark matter particles get excited and emit 3.5 keV radiation
  or electrons.
%\end{frame}
%\begin{frame}

  {\bf About the Xenon1T and DAMA-rates:}
%775
  \begin{itemize}
  \item{\color{blue} Absolute rates very crudely} Our estimate of the
    absolute rates for the
    two experiments are very very crude, because we assume that the dark matter
    particles - in our model small macroscopic systems with ten thousands
    of nuclei inside them - can have an exceedingly smooth distribution of lifetimes on a logarithmic scale.
     These calculations are discussed in section \ref{sec5}.

  \item{\color{blue} The ratio of rates} The ratio of the rates in the
    two experiments  - Xenon1T electron recoil excess  and DAMA - should in
    principle be very accurately predicted in our model, because they are supposed
    to see exactly the same effect just in two different detectors in the same underground laboratory below the Gran Sasso mountain!
    %{\color{red} We get {\bf only} order of magnitude agreement}
    One would therefore expect the rates to be the same, but the Xenon1T rate is 250 times smaller than the DAMA rate. We briefly refer to a possible resolution of this problem, which needs further study, in section \ref{sec5}.
    \end{itemize}
 % \end{frame}
\section{Impact}
%\begin{frame}
\label{Impact}
  {\bf Illustration of Interacting and Excitable Dark Matter Pearls}

  \includegraphics[scale=0.6]{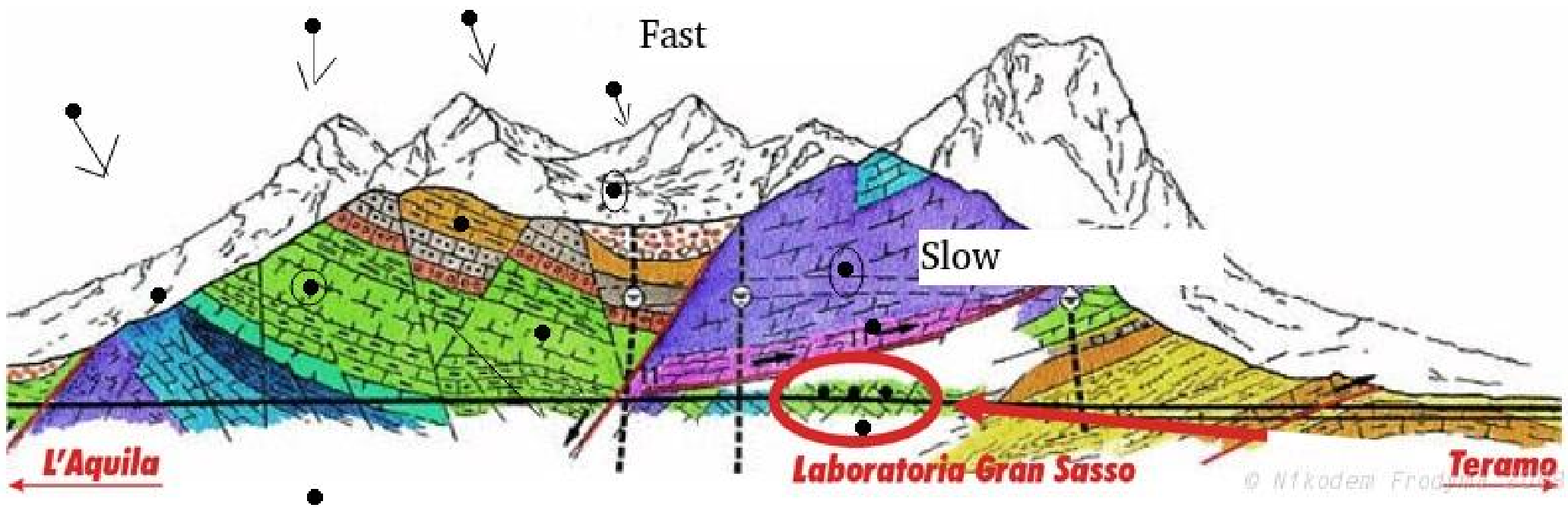}

  The dark matter pearls come in with high speed (galactic velocity), but
  get stopped down to much lower speed by interaction with the air and the shielding
  mountains, whereby they also get excited to emit 3.5 keV X-rays or
  {\em electrons}.
%\end{frame}
%\begin{frame}

  \medskip

  {\bf Pearls Stopping and getting Excited in Earth Shield}

  What happens when the dark matter pearls in our model of less than atomic
  size hit the earth shielding above the experimental halls of e.g.
  DAMA?

  \begin{itemize}
  \item{\color{blue} Stopping} Taking it that the pearls stop in
    the earth: The pearls are stopped in about $5*10^{-6}s$
    from their galactic speed of about 300 km/s down to a speed ~49 km/s
    below which
    collisions with nuclei can no longer excite the $3.5 \; keV$ excitations.
    The stopping length, modulo a logarithmic factor, is
    $\frac{1}{4}m$.

    But taking it that they stop in the air, which is more likely:
    They are stopped over a range of about 7 km - as the atmosphere
    density goes up with a factor $e=2.71..$ over such a range in about
    $2*10^{-2}s$.
  \end{itemize}
%\end{frame}
%\begin{frame}
%830
%  {\bf Pearls stopping and getting excited in Earth shileding.(continued)}

  %\vspace{-5mm}
  \begin{itemize}
  \item{\color{blue}Excitation}
    As long as the velocity is still over the ca. 49 km/s collisions with nuclei in the
    shielding can excite the electrons inside the pearl by 3.5 kev or more
    and make pairs of quasi electrons and holes say. We expect that often the
    creation of (as well as the decay of) such excitations require electrons to
    pass through a (quantum) tunnel and that consequently there will be decay half lives of
    very different sizes. We hope even up to many hours or days...
  \end{itemize}
%\end{frame}
%\begin{frame}

 % {\bf Stopping and Exciting Pearls (yet continued)}

  \begin{itemize}

  \item{\color{blue} Slowly sinking:}
    After being stopped in of the order of $\frac{1}{4}m$ of the shielding,
    the pearls continue with a much lower velocity driven by the gravitational
    attraction of the Earth. After say about 26 hours a pearl reaches the
    1400 m down to the laboratories. Most of the pearls have returned to their
    ground states, but some exceptionally long living excitations survive.
  \end{itemize}

  %{\bf \color{red}
   {\em Note that the slowly sinking velocity is so low that
    collisions with nuclei cannot give such nuclei enough speed to excite the
    scintillation counters neither in DAMA nor in Xenon-experiments.}
%\end{frame}
%\begin{frame}

  %{\bf Stopping and Exciting (yet, yet continued)}

  \begin{itemize}

  \item{\color{blue} Electron or $\gamma$ emission}
    Typically the decay of an excitation could be that a hole in the Fermi sea
    of the electron cloud of the pearl gets filled by an outside electron
    under emission of another electron by an Auger-effect. The electron must
    tunnel into the pearl center. This can make the decay lifetime
    become very long
    and very different from case to case.
    \end{itemize}
%\end{frame}

%\begin{frame}

  {\bf Emission as electrons or photons makes Xenon-experiments not see events,
    except...}

  That the decay energy is released most often as electron energy means that such events
  are discarded by most of the Xenon-experiments, which only expect the nucleus
  recoils to be dark matter events. This would explain the long standing controversy
  consisting in DAMA seeing dark matter with a much bigger rate than the upper
  limits from the other experiments.

  Rather recently though Xenon1T looked for potential excess events among the
  {\em electron recoil events} and found $16 \; events/year/tonne/keV$
  in the lowest keV-bands over a background of the order of
  $(76 \pm 2) \; events/year/tonne/keV$.

  In our model this rate should be compatible with the DAMA event rate.
  However they deviate by a factor of 250. It therefore appears that we need
  the pearls to run much faster through the xenon-apparatus than through the DAMA one.

%\end{frame}

  \section{\label{sec5} Numerical Rates for DAMA and
    Xenon1T-electron-recoil-excess}
%\label{sec5}
      \subsection{The Kinetic Energy Flux from Dark Matter}
      \label{kef}
      The dark matter density $D_{sol}$ in our part of the Milky Way and its
      velocity $v$ are of the orders of magnitude
\begin{eqnarray}
D_{sol} &=& 0.3 \; GeV/cm^3\\
%&=& 0.3GeV*1.78*10^{-27}kg/GeV*10^{6}/m^3\\
&=& 5.34 *10^{-22}kg/m^3\\
v&=& 300 \; km/s \quad \hbox{(relative to solar system)}
\end{eqnarray}
This gives a kinetic energy density
\begin{eqnarray}
D_{kin \; energy} &=& \frac{1}{2}v^2 D_{sol}\\
&=& 0.5*(10^{-3})^2c^2* 5.34*10^{-22}kg/m^3\\
%&=& 0.5 *5.34 *10^{-28}c^2kg/m^3= 0.5*4.81*10^{-11}J/m^2\\
&=&  %2.67 *10^{-28}c^2kg/m^3 =
2.40 *10^{-11}J/m^3
\end{eqnarray}
meaning an influx of kinetic energy
\begin{eqnarray}
\hbox{``power per $m^2$''}&=& vD_{kin \; energy}\\
&=& \frac{1}{2}v^3D_{sol}\\
&=& 3*10^5m/s *2.40*10^{-11}J/m^3\\
&=& 7.2*10^{-6}W/m^2
\end{eqnarray}

Distributing this energy rate over the amount of matter down to
the depth $1400 \; m$ with density $3000 \;kg/m^3$ we obtain the energy rate
per kg

\begin{eqnarray}
  \hbox{`` power to deposit''}&=& \frac{7.2*10^{-6}W/m^2}{1400 \; m * 3000 \; kg/m^3}\\
  &=& 1.7*10^{-12}W/kg.
  \end{eqnarray}

However, assuming that all the events from the
dark matter - as given by the modulated part of the signal found by
DAMA-LIBRA -
are just due to decays with the decay energy 3.5 keV, the rate of energy
deposition per kg observed by DAMA-LIBRA \cite{DAMA2} is
\begin{eqnarray}
  \hbox{``deposited rate ''} &=& 0.0412 \; cpd/kg*3.5 \; keV\\
  &=& \frac{0.0412 \; cpd/kg*3.5*1.6*10^{-16}J }{86400 \; s/day}\\
 &=& 2.7*10^{-22}W/kg,
 \end{eqnarray}
  which is
\begin{eqnarray}
\frac{2.7*10^{-22}W/m^2}{1.7*10^{-12}W/m^2}
%&=& \\
  &=& 1.6 *10^{-10} \; \hbox{times as much.}
  \end{eqnarray}

We can express this by saying that there is a need for a suppression factor
$suppression$ being $1.6*10^{-10}$ for the DAMA-LIBRA rate. For the excess
of the electronic recoil events found by Xenon1T the corresponding suppression
factor must be the 250 times smaller number. This is because the event rate
of these
excess electron recoil events is 250 times smaller than that of the modulation
part of the DAMA rate and the depth of the experiment under the earth is the
same 1400 m. Thus we summarize the {\em experimentally} determined suppression
factors:
\begin{eqnarray}
  suppression_{DAMA} &=& 1.6*10^{-10}\\
  suppression_{Xenon1T}&=& \frac{1.6*10^{-10}}{250}=  6.4*10^{-13}.
\end{eqnarray}
%982
\subsection{Estimating ``suppression'' theoretically}

The idea for obtaining theoretical estimates of these suppression factors
is to say that the observed events come from excitations of our pearls with a
lifetime of the order of the time it takes for the pearl, after its excitation
under its stopping in the air or in the stone above the experiments, to reach
down to the experimental detectors.  We here assume the scattering cross section of dark matter on
ordinary matter to be similar to that on dark matter.  So we estimate the passage time
%giving the lifetime needed for the survival of the excitation
of the pearl down
to the detectors as being of the order of 26 hours, by using the
low velocity value for the cross section over mass ratio
\begin{eqnarray}
  \hbox{To be used for passage velocity: }\frac{\sigma}{M}&=& 150 \; cm^2/g
\end{eqnarray}

Once the pearl has been stopped so much that its velocity is only upheld
by the gravitational
field with the acceleration $g=9.8 \; m/s^2$, the terminal velocity will
be obtained formally from the drag-equation\footnote{Strictly speaking this equation
is only valid if the pearl velocity is greater than the thermal velocity of the nuclei
in the shielding and so needs further study.}
\begin{eqnarray}
  \hbox{Drag force} \;
  D
=  gM &=&0.5*C_d*\sigma*\rho v^2.
\end{eqnarray}
Here $\rho$ is the density of the material passed through and the drag coefficient
$C_d$ is of order unity (so the 0.5 is hardly relevant). That is to say
the terminal velocity becomes:
\begin{eqnarray}
  v_{terminal}&\approx& \sqrt{\frac{g}{\frac{\sigma}{M}*\rho}}\\
  &\approx&\sqrt{\frac{9.8 \; m/s^2}{150 \; cm^2/g*3 \; g/cm^3}}\\
    &=&\sqrt{2.2 \; cm^2/s^2} =1.5 \; cm/s,
\end{eqnarray}
which allows a pearl to pass though 1400 m in
\begin{eqnarray}
  \hbox{``passage time'' }&=&\frac{140000 \; cm}{1.5 \; cm/s}\\
  &=& 93000 \; s=26 \; hours.
\end{eqnarray}

\subsection{Equally hard to excite and to de-excite}

In order that there can be any de-excitations of the pearls after such 26 hours
it is of course needed that an appreciable part of the possible 3.5 keV excitations
of our pearls have lifetimes of this order of magnitude.
A priori these excitations are excitons for which the electron and hole can
be close by and decay rapidly or it is possible that one of the partners is
outside in the electron cloud and long lived.
By arguing that some tunnelling of electrons in or out or around in the
pearl may be needed for some (de-)excitations, we can claim that
the lifetimes for the various excitation possibilities are smoothly distributed
over a wide range in the logarithm of the lifetime;
%a distribution
%of lifetimes for the various excitation possibilities being distributed
%in the logarithm of the lifetime with a rather smooth distribution of wide
%range;
then there will be some pearl-excitations with the appropriate
lifetime, although somewhat suppressed by a factor of the order of
$1/width$ where the $width$ here is the width of the logarithmic distribution.
We shall take this $width$ to be of order $\ln suppression_{DAMA} \sim 23$. But more importantly:
If a certain excitation is long-lived, it is also hard to produce.
So we shall talk about an effective `` stopping'' or ``filling time''
for a pearl passing into the Earth, and imagine that during this
``stopping'' or ``filling time'' the excitations of the pearls
have to be created. So the probability for excitation or
$suppression$ would be expected to be
\begin{eqnarray}
  suppression &\approx& \frac{\hbox{``filling time''}}{\hbox{``lifetime''}}.
\end{eqnarray}
   If the excitation happens to be
of sufficiently long lifetime - say of order 26 hours - then we can expect it
to have a sensible chance of de-exciting just in the experimental
detectors in Gran Sasso, DAMA or Xenon1T say.
% if the ``lifetime'' is of order 26 hours.

But what shall we take for this ``stopping'' or ``filling time''?
A relatively simple idea, which is presumably right, is to say that
the stopping takes place high in the atmosphere
because a pearl entering the Earth's atmosphere with galactic speed
will be slowed down
in the high air with a $\frac{\sigma}{M} \sim 2 \; cm^2/g$.
Now the density of the atmosphere rises by a factor $e=2.718...$
per about 7 km. So as the slowing down begins it will, because of this rising
density, essentially stop again after 7 km. Thus the time during which
the pearl is truly slowing down in speed and forming 3.5 keV excitations is of the order of the time
it takes for it to run 7 km. With the pearl velocity of about $300 \; km/s$ (essentially the
escape velocity for the galaxy) we then have
\begin{eqnarray}
  \hbox{``stopping time''}&\approx& \frac{7 \; km}{ 300 \; km/s}\\
  &=& 0.023 \; s
    \end{eqnarray}

The crucial factor, which we believe to be most important, is that in
order to excite an excitation with a lifetime of the order 93000 s
it would a priori need 93000 s so that, if we only have 0.023 s, then
there will be a suppression:
\begin{eqnarray}
   suppression &=&
  \frac{\hbox{``stopping time''}}{\hbox{``lifetime''}}\\
    &\approx&  \frac{\hbox{``stopping time''}}{\hbox{``passage time''}}\\
      &\approx& \frac{0.023 \;s}{93000 \; s}\\
      &=& 2.5*10^{-7}.
\end{eqnarray}

This crudest estimate has to be compared with the experimental suppressions
given above
\begin{eqnarray}
  suppression_{DAMA} &=& 1.6*10^{-10}\\
  \Rightarrow \frac{suppression_{theory}}{suppression_{DAMA}}&=&\frac{2.5*10^{-7}}
              {1.6*10^{-10}} \\
              &=& 1.6*10^3\\
              suppression_{Xenon1T}&=& \frac{1.6*10^{-10}}{250}=  6.4*10^{-13}\\
              \Rightarrow \frac{suppression_{theory}}{suppression_{Xenon1T}}
              &=& \frac{2.5*10^{-7}}{6.4*10^{-13}}\\
              &=& 3.9 *10^5.
\end{eqnarray}

But here can be several corrections to $suppression_{theory}$, at least we should correct by the
width in logarithm of the supposed distribution of the lifetimes  among the
different excitations. Above we suggested a factor 23, which would bring the
DAMA rate to only deviate by about a factor 100. Our estimate is of course
extremely uncertain.

We can never get the DAMA rate and the electron recoil excess rate from Xenon1T agree
with the same estimate in as far as they deviate by a factor 250. Our only
chance is in a later paper to justify say the story that, because the
scintillator in which the Xenon1T events are observed is {\em fluid} while the
NaI in the Dama experiment is solid, the pearls pass much faster through
the Xenon1T apparatus than they pass through the DAMA instrument. Imagine say
that the pearls partly hang and get stuck in the DAMA experiment, but that
they cannot avoid flowing down all the time while they are in the fluid Xe
in the Xenon1T scintillator.

\section{3.5keV}
%\begin{frame}
\label{sec7}
  { Order of magnitudewise we see 3.5 keV in {\huge 3} different places.}

%\begin{figure}
  \includegraphics[scale=0.8]{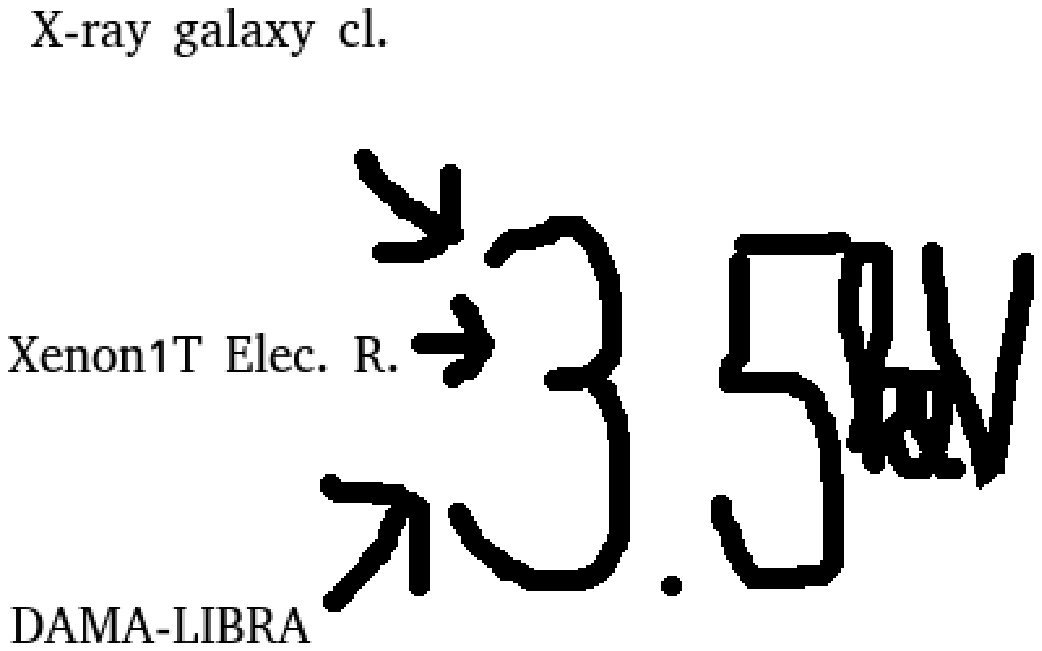}

%\end{frame}
%\begin{frame}

 %\caption{\label{coincidence} The 3.5 keV crude Coincidence}
%\end{figure}

 % \vspace{-3mm}
  The energy level difference of about 3.5 keV occurring in 3 different
  places is important evidence motivating our
  model of dark matter particles being excitable by 3.5 keV:
  \begin{itemize}
  \item{\color{blue} The line} From places in outer space with a lot of dark
    matter, galaxy clusters, Andromeda and the Milky Way Center, an unexpected
    X-ray line with photon energies of 3.5 keV (to be corrected for Hubble
    expansion...) was seen.
  \item{\color{blue} Xenon1T} The Dark matter experiment Xenon1T did not find
    the standard nuclei-recoil dark matter, but found an excess of {\em electron-recoil}
    events with energies concentrated crudely around 3.5 keV.
    %in the low end where it gets impossible to measure
    %below 2 keV, in fact with concentration crudely around the 3.5 keV.
  \item{\color{blue} DAMA} The seasonally varying component of their events lie in energy
    between 2 keV and 6 keV, not far from centering around 3.5 keV.
    \end{itemize}
%\end{frame}

%\begin{frame}

 % {\bf Decay Observation in DAMA and Xenon1T almost a Must}

  We take it seriously and not as an accident that both DAMA and Xenon1T see
  events with energies of the order of the controversial astronomical 3.5 keV X-ray
  line. We are thereby driven towards the hypothesis that the energies
  for the events in these underground experiments are
  determined from a decay
  of an excited particle, rather than from a collision with a particle in the
  scintillator material. It would namely be a pure accident, if a collision
  energy should just coincide with the dark matter excitation energy
  observed astronomically.

  {\em So we ought to have decays rather than collisions!}
%\end{frame}
%\begin{frame}
  {\em How then can the dark matter particles get excited ?}

  You can think of the dark matter pearls in our model hitting electrons
  and/or nuclei on their way into the shielding:
  \begin{itemize}
  \item{\color{blue} Electrons} Electrons moving with the speed of the dark
    matter of the order of 300 km/s toward the pearls in the pearl frame
    will have kinetic energy of the order
    \begin{eqnarray}
      E_e &\approx & \frac{1}{2}*0.5 \; MeV*(\frac{300 \; km/s}{3*10^5km/s})^2=
      0.25 \; eV.
    \end{eqnarray}

  \item{\color{blue}Nuclei} If the nuclei are say Si, the energy in the collision
    will be 28*1900 times larger $\sim 5*10^4$ * $0.25 \; eV$
    $\approx$ 10 keV. That would allow a 3.5 keV excitation.

    To deliver such $\approx$ 10 keV energy the nucleus should hit
    something harder than just an electron inside the pearls. It should
    preferably hit a nucleus, e.g. C, inside the pearl.
    \end{itemize}

%\end{frame}

\section{Conclusion}
\label{Conclusion}
%\begin{frame}

 % {\bf Conclusion, Model description}

  \begin{itemize}
  \item { We have described a seemingly viable model for dark
    matter consisting of atomic size but macroscopic pearls. These pearls consist of a
    bubble of a new speculated type of vacuum containing some normal
    material - presumably carbon - under the high pressure of the skin
    (surface tension). They each contain about a hundred  thousand  nucleons in the
    bubble of radius about $5*10^{-12}m$.}
  \item { The electrons in a pearl have partly been pushed out of the genuine
    bubble of the new vacuum phase, out to a distance of about $5*10^{-11}m$.}
    %\end{itemize}

%\end{frame}

%\begin{frame}

 % {\bf Conclusion on Observations considered relative to the Model}

  We have compared the model or attempted to fit:

  %\begin{itemize}

  \item { Astronomical suggestions for the self interaction of dark matter
    in addition to pure gravity.}

  \item {The astronomical 3.5 keV X-ray emission line found by satellites, supposedly from dark
    matter.}

  \item {The underground dark matter searches.}

    \end{itemize}

%\end{frame}
%\begin{frame}

  { We list below the quantities we have crudely estimated:}
  \begin{enumerate}
  %\begin{itemize}
  %\item{\color {red} 1.}
  \item The low velocity cross section divided by mass.
  %\item{\color{red} 2.}
  \item That the signal from Xenon1T and Dama should agree
    %(except for different scintillation efficiencies and)
    except that the pearls may
    run with different velocities through the scintillator materials, because
    the xenon-instruments use {\em fluid} xenon, while the DAMA-LIBRA
    experiment uses the solid NaI.
  %\item{\color{red} 3.}
  \item The absolute rate of the two underground experiments.
    (But unfortunately unless we explain the ratio of the rates for the
    two experiments as say due to the different velocities through the scintillator materials,
    we can of course never predict the absolute
    rate to be better than deviating by about a factor of 250 with
    at least one of them.)
  %\item{\color{red} 4.}
  \item The rate of emission of the 3.5 keV X-ray line from the Tycho
  supernova remnant due to the excitation of our pearls by cosmic rays \cite{Bled20}.
  %the Jeltema et al. Tycho SN remnant
  %  3.5 keV
  %\item{\color{red}5.}
  \item Relation between the frequency 3.5 keV and the overall emission
    rate of this X-ray line observed from galaxy clusters etc.
  \item We also previously predicted the ratio of dark matter to
    atomic matter (=``usual'' matter) in the Universe to be of order 5
    by consideration of the binding energies per nucleon in helium and heavier nuclei,
    %from some simple binding energies of nuclei,
    assuming that the atomic matter at some
    time about 1 s after the Big Bang was spit out from the pearls under a fusion explosion
    from He fusing into say C \cite{Dark1}.
   %\end{itemize}
   \end{enumerate}
 % \end{frame}

  \subsection{Outlook}

  At the end we want to mention a few ideas which we hope will be developed
  %of the works we hope that we or others will work on in
  as a continuation of the present model:
\begin{itemize}
\item{\bf Speculative Phase from QCD.}
  QCD and even more QCD with  Nambu-Jona-Lasinio type spontaneous symmetry breaking
  is sufficiently complicated, that possibly a new phase appropriate for our pearls could be hiding there.
  There is already an extremely interesting observation \cite{KKS}.

\item{\bf Relative Rates of DAMA and Xenon1T.} A crucial test for our model is
  to reproduce the relative event rates in DAMA and in the excess
  of electron recoils in Xenon1T. This requires a careful study of the viscosity
  of fluid xenon and the properties of our pearls.
  %For the working of our picture
  %it is crucial and a test which needs rather high accuracy to see if
  %the viscosities of the fluid xenon and the properties of our pearls
  %can accommodate for the right relative rates in DAMA and in the excess
  %of electron recoils in Xenon1T.

\item{\bf Walls in the Cosmos.} With the usual expectations for the density per
  area or equivalently tension $S$, cosmology would be so severely
  changed by such domain walls that models with say $S^{1/3}\ge 10$ MeV
  are phenomenologically not tenable. However, with our fit to a surprisingly
  small tension with $S^{1/3}\approx 2.2$ MeV it just barely
  becomes
  possible to have astronomically extended domain walls, e.g. walls around the
  large voids between the bands of galaxies; so that these voids could be say
  formally huge dark matter pearls, though with much smaller density.
  In fact a series of domain walls with our fitted $S=2.2^3$ MeV$^3$ with
  distances
  between them of the order of 13 milliard light years would have a
  density not much different for that of the universe we know.

\item{\bf New Experiments?} According to our estimates the observed rate of
  decays of our dark matter pearls should be larger the less shielding
  they pass through. So an obvious test of our model would be to make a DAMA-like
  experiment closer to the earth surface where we would expect a larger
  absolute rate than in DAMA, although there might of course be more background.
  Actually such an experiment is already being performed by the ANAIS group \cite{ANAIS},
  but they have so far failed to see an annual modulation in their event rate.

\end{itemize}

\section*{Acknowledgement}
HBN thanks the Niels Bohr Institute for his stay there as emeritus.
CDF thanks Glasgow University and the Niels Bohr Institute for hospitality and support.
Also we want to thank many colleagues for discussions and for organizing
conferences, where we have discussed previous versions of the present model,
especially the Bled and Corfu meetings.

\

\end{document}